\documentclass[12pt,a4paper]{article}
\usepackage{geometry}
\usepackage[round]{natbib}
\usepackage{mathtools}
\usepackage{amsfonts}
\usepackage{amssymb}
\usepackage{graphicx}
\usepackage{placeins}
\usepackage[labelfont=bf]{caption}
\usepackage[skip=-0.25\textwidth]{subcaption}
\usepackage{enumitem}%to modify item bullets
\usepackage{titlesec}
\usepackage[titletoc,title]{appendix}
\usepackage{xpatch}
\xpretocmd{\eqref}{equation\,}{}{}
\usepackage{authblk}
\title{\textbf{Auto-electrophoresis in non-Newtonian media: interaction of rheology and electrocatalytic parameters}}
\author[1]{Antarip Poddar}
\author[1]{Aditya Bandopadhyay\thanks{Email: aditya@mech.iitkgp.ernet.in}}
\author[1]{Suman Chakraborty\thanks{Email: suman@mech.iitkgp.ernet.in}}
\affil[1]{Department of Mechanical Engineering, Indian Institute of Technology Kharagpur, Kharagpur, West Bengal - 721302, India}
\date{}                     
\begin{document}
	\maketitle
	\begin{abstract}
		\noindent
		Recent findings of possible applications of bio-friendly synthetic self-phoretic swimmers,  have motivated the researchers in investigating the various motion-generating mechanisms to optimize the operating characteristics of the same. In this paper we model the auto-electrophoretic motion of a bimetallic $ \mathrm{(Au-Pt)} $ spherical swimmer in a non-Newtonian medium. In view of the fact that many bio-fluids closely follow the generalized Newtonian rheology, the rheology of the surrounding swimming medium is considered to follow the Carreau-Yasuda viscosity model. Further to capture the experimentally observed effects of peroxide concentration and feedback of generated proton concentration on the surface cation flux, we incorporate a Michaelis-Menten like surface reaction kinetics.  Subsequently the equations are numerically  solved and the results are verified in the limiting condition of a Newtonian medium.  The electrocatalytic efficiency defined as a ratio between the cost of swimming of an equivalent passively dragged particle and the input chemical energy via surface reaction, is utilized to assess the influence of the rheological parameters on the swimmer performance. The results indicate  that the shear thinning and shear thickening nature of the surrounding fluid causes an enhancement and reduction in swimmer velocity, respectively.  However, an increase in the shear thinning effect does not always guarantee an efficient operation of the microswimmer. A competition between the particle velocity and reduction in the drag force on the particle, decides whether the propulsion efficiency will have a relative augmentation or attenuation with varying power law index. We further report the existence of maximum efficiency points for some specific range of Weissenberg number and optimum power law indices. Moreover the behaviour of such optimum operating conditions strongly depends on the non-trivial and highly coupled interplay among the rheology and electrocatalytic parameters. In addition, the incorporation of a detail surface kinetics model enabled us to demonstrate an experimentally consistent size-dependency of the swimmer velocity. The significant enhancement in electrocatalytic swimming efficiency as compared to its Newtonian medium counterpart and its remarkable alterations with the  electrocatalytic parameters, provide a new direction in choosing the optimum conditions for a favourable performance of synthetic microswimmers. 
	\end{abstract}
	\section{Introduction}
	Inspired by the self-generated propulsion phenomena in biological microorganisms, scientists have successfully mimicked synthetic micro and nano-motors showing autonomous movements powered by various autophoretic mechanisms. The ability of these artifical swimmers in consuming either an externally introduced chemical or a locally available enzyme \citep{Dey2015} as a fuel source and no external mechanical energy supply, have made them suitable for considerations in a wide range of practical applications including transport of cargo, targetted drug delivery \citep{Patra2013}, biosensing, diagnosis  \citep{Chalupniak2015}, environmental remediation \citep{Katuri2016} and many others \citep{Dey2016,Peng2017,Duan2015,Jurado-Sanchez2017,Ebbens2016}. 
	
	Among the various extensively studied  self-propulsion mechanisms the auto-electrophoresis (AEP) has received a significant attention among the researchers owing to its agreement with vast number of experimental observations \citep{Brown2014}. In this regard, an autonomous motion of gold-platinum bimetallic swimmers were first observed in the presence of hydrogen peroxide solution as fuel source \citep{Paxton2004,Paxton2006}. Subsequently various other metallic combinations and carbon fibers, were also found to produce effective swimming \citep{Fournier-Bidoz2005,Wang2006,Mano2005}. Parallelly consistent theoretical models were proposed to have a deep understanding of the intricate physical mechanisms behind the effective conversion of chemical to useful mechanical energy in these miniaturized devices, mainly  cylindrical or spherical in shape \citep{Moran2010,Moran2011,Sabass2012a,Ibrahim2017,Brown2017}. An efficient conversion of chemical to useful mechanical energy still remains a challenge owing to a dominant viscous friction at the micro-scale.
	{In their study \citet{Kreissl2016} showed that the electrocatalytic swimming efficiency can be maximized if the surface reactivity cites can be designed to be concentrated near the polar regions of a spherical swimmer. Similarly \citet{Nourhani2015} found that incorporation of an inert region in between the active polar regions, enhances the capability to harness more mechanical power for a fixed reaction flux. However, these studies suggest the requirement of a precise fabrication technique capable of creating a predetermined distribution of the reactive sites.}
	
	Although the toxicity of hydrogen peroxide was previously thought to be a barrier in employing bimetallic microswimmers in biological environments,  later  the researchers have  utilized bio-friendly fuel sources such as glucose, methanol, urea or other locally present enzymes \citep{Kumar2013,Yoshizumi2013,Schattling2015,Pavel2014,Dey2015}. Thus the bimetallic microswimmers have emerged with a potential of being used in variety of in vitro and in vivo applications \citep{Wang2012,Guix2014}. With growing interest in employing the synthetic microswimmers in biological systems, it is of utmost importance to access their motion characteristics and performance in complex bio-fluid mediums \citep{Neves2012,Chalupniak2015}. There exists a spectrum of bio-fluids whose rheological behaviour follows a shear-rate dependent viscosity well described by a generalized Newtonian model (e.g. human blood \citep{Walburn1976,Chakraborty2005}, synovial fluid \citep{Kren2007}, physiological mucus \citep{Hwang1969,Davies2015}, vaginal gels \citep{Yu2011}).  In order to capture the consequences of  complex behaviour of the  swimming medium in the autonomous movement of artificial microswimmers, we choose the Carrea-Yasuda model which has the ability to predict the Newtonian as well as power law behaviour of bio-fluids in different limiting conditions. Moreover, the chosen C-Y model has been previously proved to be suitable for  modeling of  swimming microorganisms \citep{Li2015,Montenegro-Johnson2013,Velez-Cordero2013,Gagnon2014,Gagnon2016} since it can successfully predict realistic finite shear stress values at both zero and infinite shear rates. The existing works in literature focusing on the squirming motion of active particles in shear thinning fluids  \citep{Datt2015,Nganguia2017}, have only taken into account a prescribed surface flow distribution and no attempt was  made to delineate the combined interplay of the motion generating mechanism and fluid rheology.
	
	In conventional electrophoresis where an external electric field is applied, the non- Newtonian fluid characteristics of the Carreau type fluid has been found to get coupled with electrokinetics and it non-trivially alters the flow field as well as particle velocity \citep{Khair2012,Lee2005}. Similar effects were also observed for electroosmosis \citep{Zhao2011}. In stark contrast, in the present case of self-electrophoresis, the electric field is endogenously created due to a heterogeneous electrocatalytic reaction at the swimmer surface. This self-generated electric field provides the required driving force for swimming. 
	
	Despite great advancements in the swimming characterization of auto-electrophoretic particle in Newtonian media, till now no attention has been directed to investigate the coupled effects of electrocatalysis phenomena and complex fluid rheology on the swimmer velocity and an efficient utilization of the fuel source. In the present work we consider the self-phoretic motion of a spherical microswimmer which has two distinct metallic faces one of which catalyzes the dissociation rate of the surrounding $ \mathrm{H_2O_2} $ solution. The heterogeneous ionic fluxes at the swimmer surface creates an asymmetry in charge distribution along the longitudinal direction, resulting in a autonomous movement of the swimmer. Consideration of a Michaelis-Menten like surface kinetics helps us to capture the effects of fuel concentration and feedback of genrated  distribution of active species on the surface cation flux. By incorporating a Carreau-Yasuda  rheology model, we capture the coupled effects of various electrocatalytic parameters and complex fluid rheology on the resulting particle motion. The complete set of dimensionless governing equations and boundary conditions are then solved numerically. Further we attempt to assess the performance of the autoelectrophoretis mechanism in generating an efficient swimming under various operating conditions.
	
	\section{Mathematical formulation}
	\label{sec:mathform}
	\subsection{Physical system}
	\label{ssec:system}
	
\begin{figure}[!htbp]
		\centering
		%		\vspace{-5ex}
		\hspace{-2ex}
		\includegraphics[width=01\textwidth]{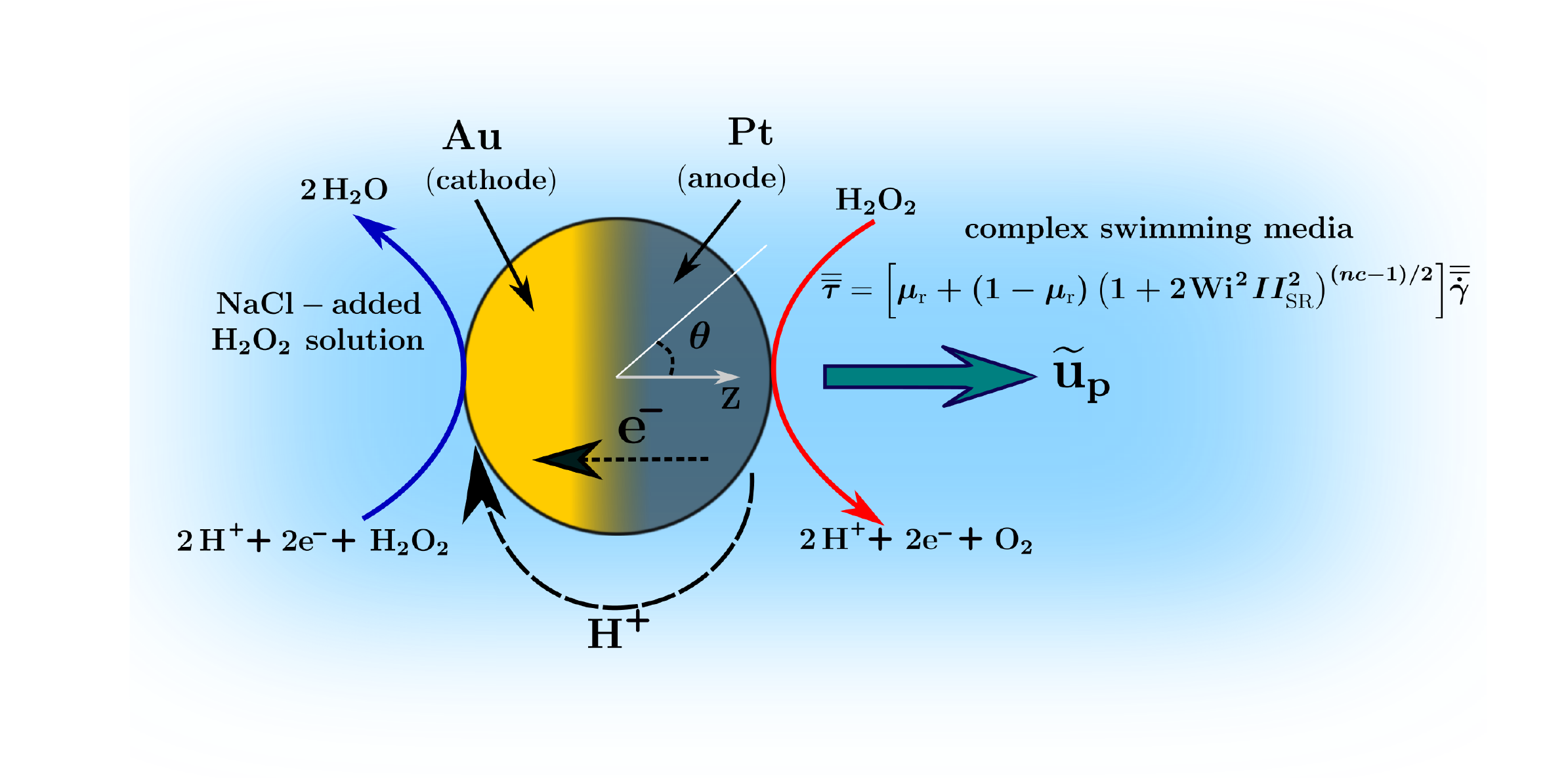}
		%	\vspace{-90ex} 
		\caption{ Schematic representation of a spherical bimetallic $\mathrm{(Au-Pt)} $ swimmer. The particle is in a salt added $\mathrm{H_2O_2}$ solution. Oxidation of $\mathrm{H_2O_2}$ takes place at the Pt-face, while the same is reduced at the Au-face. The gradual variation of colour of the swimmer surface pictorially represents a smooth variation of the reaction rate from the platinum to gold side.  The surrounding complex swimming medium obeys the Carreau-Yasuda constitutive model for fluid rheology. The heterogeneous reaction kinetics at the surface creates an endogenously generated proton current which in turn creates a body force for the ionic species already present in the solution. A co-moving  axisymmetric coordinate system is  fitted at the particle center. The particle shows a swimming velocity $ \widetilde{u}_\mathrm{p}$  along the +ve $ z $  axis of the chosen coordinate system. }
		\label{fig:schematic}
	\end{figure}

	We consider a physical situation (as shown in figure \ref{fig:schematic}) where a $ \mathrm{(Au-Pt)} $ bimetallic microswimmer of spherical shape is suspended in an aqueous salt solution with a fraction of hydrogen peroxide. Due to its immediate contact with an electrolyte, the swimmer acquires a net surface charge. Subsequently to maintain an overall charge neutrality of the system, a diffusive screening layer of counterions is formed adjacent to the particle surface. Such an arrangement of ions is widely known as the electrical double layer or EDL. Consequently the catalytic segment ($\mathrm{Pt-} $ face) facilitates a spontaneous dissociation of hydrogen peroxide molecules into oxygen and water. This is known to occur in two successive steps - oxidation of $ \mathrm{H_2 O_2} $ at the anode and its reduction at the cathode $ \mathrm{(Au-face)} $ \citep{Moran2017}. The resulting high concentration of protons relative to the cathode segment, creates a gradient of proton density, thus establishing an asymmetric charge distribution around the particle so that electromigration balances the induced proton flux. Due to this phenomenon, even in the absence of any externally applied electric field an electrochemically-induced electric field comes into existence. This further creates a body force that drives the ions in the diffusive screening layer. Owing to the migration of protons towards  the cathodic segment of the swimmer, the body force on the surrounding fluid is generated from the anode to cathode. This causes the particle to move (with respect to the inertial reference frame) in the opposite direction with the anodic metal face in the front side.        
	
	Although demonstrated for a direct $ \mathrm{H_2O_2} $ fuel, the present study can be easily extended for indirect generation of $ \mathrm{H_2 O_2} $ from glucose catalyzed by an immobilized coating of glucose oxidase $ (\mathrm{GO_x}) $ on the particle surface \citep{Kumar2013}. Similar other bimetallic combinations such as $ \mathrm{Cu-Pt\, ,Ag-Pt\, ,Zn-Pt} $ can be used in place of the $ \mathrm{(Au-Pt)} $ electrode pair. 
	
	We adopt a co-ordinate system fitted at the center of the particle (please refer to figure \ref{fig:schematic}) and consider that the reaction rate varies along the polar direction $ (\theta) $ along the surface of the particle. The length scale is chosen as the particle radius $ a $ and viscosity values are non-dimensionalized with the zero shear rate viscosity in the presently chosen Carreau-Yasuda viscosity model $ (\widetilde{\mu}_0) $. The bulk concentration of the proton $ (\mathrm{H^+}) $ and hydroxide ions $ (\mathrm{OH^-}) $ are considered to be equal $\widetilde{c}^+_\mathrm{H}=\widetilde{c}^-_\mathrm{OH} $. In addition we consider a symmetric electrolytes with valency=1 (eg. NaCl, KCl) is employed and  consequently assume equal bulk concentration of the salt ions of opposite sign $ (\widetilde{c}_0 )$.

	\subsection{Governing equations and boundary conditions}
	\label{ssec:math_gov_eq}
	
	In order to identify the key dimensionless variables required to obtain a compact description of the problem, the reference scales for various variables are taken as:  length $ \sim a$ (particle diameter), electrostatic potential  $ \sim k_B T/e$ (thermal voltage), ionic concentration $ \sim \widetilde{c}_0$ (bulk salt ion concentration) and viscosity $ \sim \widetilde{\mu}_0$ (constant viscosity  at zero shear rate in Carreau-Yasuda model). Thus the naturally evolving dimensional references for velocity and reaction coefficients turn out to be $(u_{ref}=\epsilon_0 \, \epsilon_r \, k_B^2 \, T^2/e^2 \, \widetilde{\mu}_0 \,a)$ and $(\widetilde{c}_0 \, D/a)$, respectively. Here onwards we drop the ${\; \widetilde{} \;} $ from the dimensional variables and subsequently derive the governing differential equations and boundary conditions in terms of the non-dimensional variables only.

	In deriving the governing differential equations for particle motion, we will assume a quasi-steady-state condition and neglect any rotational diffusion associated with the Brownian effects. \citet{Wheat2010} argued that the rotational diffusivity of the spherical swimmers of radius $ a \gtrsim 1 \, \mathrm{\mu m} $, is significantly lower than other commonly used bimetallic swimmer geometries like a cylindrical-shaped one. Along similar lines their experimental observations also showed a pronounced unidirectional motion of the spherical bimetallic swimmers. Moreover the relative importance of the rotational and directional motion can be estimated by considering the important time scales of important mechanisms. The rotational diffusion time scale can be approximated as $ t_\mathrm{rot} \sim \dfrac{1}{D_\theta}=\dfrac{\widetilde{\mu}_0 \, a^3}{k_B \, T} $, where $ D_\theta $ denotes the rotational diffusivity. The other important time scale associated with the spatial adjustment of the ionic concentration is given by $ t_\mathrm{D} \sim \dfrac{a^2}{D} $, where $ D $ is the ionic diffusion coefficient. A comparison shows $ \dfrac{t_\mathrm{D}}{t_\mathrm{rot}} \sim \dfrac{k_B\, T}{D \, \widetilde{\mu}_0 \, a} \ll 1 $. Thus the present analysis remains applicable for the instantaneous velocity of swimmers having a length scale in the micrometer range \citep{Golestanian2007}.

	The conservation of charges is satisfied by the Poisson equation given by 
	\begin{equation}\label{eq:poisson}
		\nabla^2\psi=-\dfrac{\rho_e}{2\,\lambda^2},
	\end{equation}
	where $ \rho_e(=\displaystyle\sum_i^{} z_i\,c_i )$ is the total charge density and $ \lambda $ is defined as the dimensionless Debye layer thickness, $ \lambda=\sqrt{\dfrac{\epsilon_0\epsilon_rk_BT}{8\pi e^2 N_A c_0\,a^2}} $.
{Keeping in view of the very thin Stern layer thickness (in the order of  \r A ) \citep{Hunter2013,Bazant2005,Brown2016}, we neglect the potential  drop across the Stern layer and assume that the potential at the particle surface $(V_S)$ effectively represents the zeta potential ($ \zeta $) at outer Helmholtz plane of the electrical double layer (EDL).} Thus the electrostatic potential satisfies the following condition at the particle surface 
	\begin{equation}\label{eq:zeta-bc}
		\psi(1,\theta)=V_S.
	\end{equation} 
	On the other hand $ \psi $ becomes uniform far from the particle surface, thus allowing us to impose the following boundary condition at the far-field boundaries:
	\begin{equation}\label{eq:bc-psi-zero}
		\psi(r\to \infty,\theta)=0.
	\end{equation}
	This serves as a reference value of the electrostatic potential in the system.
	
	The ionic fluxes are given as
	
	\begin{equation}\label{eq:ion-flux}
		\mathbf{j}_i=-\nabla c_i -z_i c_i \nabla \psi + \textrm{Pe}\, \mathbf{u} \, c_i,
	\end{equation}
	where $ z_i $ is the valency of the respective ionic species,  $ \mathrm{i=H^+, OH^-,Na^+,Cl^-} $; $ c_i $ is the ionic concentration and $ \psi $ denotes the  electrostatic potential.  With the present assumption of steady, the conservation of ionic species are guaranteed by the equations: 
	\begin{equation}\label{eq:NP-bal}
		\nabla \cdot \mathbf{j}_i =0.
	\end{equation}
	Now far from the particle surface $ (r \to \infty) $ the ionic species reach the bulk concentrations, given as:
	\begin{equation}\label{eq:bulk-conc-bc}
		\begin{aligned}
			& \text{for }\,\mathrm{Na^+, Cl^-}: && c_i=1 \\
			& \text{for }\,\mathrm{H^+, OH^-}:  && c_i=\delta,  
		\end{aligned}
	\end{equation} 
	where $ \delta $ is the dimensionless bulk concentration of H$^+ $ or  OH$^- $ ions, i.e. $ \delta= \widetilde{c}^+_{\mathrm{H},0}/\widetilde{c}_0 =\widetilde{c}^-_{\mathrm{OH},0}/\widetilde{c}_0 $.
	
	At the particle surface $ (r=1) $ the normal ionic fluxes satisfy:
	\begin{equation}\label{eq:normal-flux-bc}
		\begin{aligned}
			& \text{for active species}\,\mathrm{(i=H^+)}: && \int_{S}^{}{} {\mathbf{\hat n}} \cdot \mathbf{j}\;  dS=\mathcal{J}(\theta) \\
			& \text{for inactive species}\,\mathrm{(i=OH^-,Na^+,Cl^-)}:  && \int_{S}^{}{} {\mathbf{\hat n}} \cdot \mathbf{j}\;  dS=0.
		\end{aligned}
	\end{equation}
	Finally, the rate of proton exchange at the swimmer surface has to consistently satisfy the following condition of net zero electric current across the generating and consuming portions: 
	\begin{equation}\label{eq:zero-net-flux}
		\int_{S}^{}{} \mathcal{J}(\theta) \;  dS=0
	\end{equation}
	
	The proper form of the surface cation-flux $ \mathcal{J}(\theta) $ appearing in \eqref{eq:normal-flux-bc}, has to be decided based on the nature of the heterogeneous surface reaction occurring at the swimmer surface. However, the exact form of $ \mathcal{J}(\theta)$ is not well established \citep{Moran2017} and the researchers have adopted various functional forms for rod-shaped \citep{Moran2011} as well as spherical microswimmers \citep{Sabass2012a,Kreissl2016,Nourhani2015,Yariv2010}, to make theoretical progress. In the present study we adopt the model proposed by \citet{Sabass2012a} for spherical swimmers based on Michaelis-Menten like reaction kinetics. This has the merit to capture the dependence of the surface cation-flux on the  $ \mathrm{H_2O_2} $ concentration  for $ c_\textrm{hp} \lesssim 5 \% $ as well as on the local ionic concentration of active species $ (\mathrm{H^+})$, where the term $ c_\textrm{hp} $ represents the  concentration of $ \mathrm{H_2O_2} $ in a dimensionless fraction of   wt/vol $(\%)$. It also  models a smooth variation of oxidation and reduction rates over the swimmer surface. Together with the consideration of negligible stern layer conductance the expression of surface cation-flux can now be written as:
	\begin{equation}\label{eq:cation-flux-equation}
		\mathcal{J}(\theta)=\underbrace{-(\kappa_r +K_r\,\cos(\theta))\, c^+_\textrm{H}\,c_\textrm{hp}}_{\textrm{models reduction}}+ \underbrace{(\kappa_o -K_o\,\cos(\theta))\,c_\textrm{hp}}_{\textrm{models oxidation}},
	\end{equation}
	where $ K_o \le \kappa_o $  and $ K_r \le \kappa_r $. 
	In deriving the above equation we have assumed same ionic diffusion coefficients $ D_i $ for all the species similar to \citet{Sabass2012a}.

	In the creeping flow limit $ (Re \ll 1) $, the linear momentum equation for fluid flow takes a form: 
	\begin{equation}\label{eq:Cauchy}
		-\nabla p\,+\, \nabla \cdot \boldsymbol{\overline{\overline{\tau}}}-\! \! \! \! \! \! \underbrace{b\, \rho_e \nabla \psi}_{\substack{\text{Self-generated} \\ \text{electrical body force}}} \! \! \! \! \! \! \!=0,
	\end{equation}
	where $ b $ is a dimensionless parameter defined as $ b=\dfrac{a\, \widetilde{c}_0 N_A \, k_B T}{\widetilde{\mu}_0 \, u_{ref}} $, which quantifies the importance of electrical body force to viscous body force. 
	The viscous stress tensor $ \boldsymbol{\overline{\overline{\tau}}} $ is related to the strain-rate tensor $  \boldsymbol{\overline{\overline{\dot{\gamma}}}}$ as 
	$ 
	\label{eq:stress-strain}
	\boldsymbol{\overline{\overline{\tau}}}=2\, \mu \boldsymbol{\overline{\overline{\dot{\gamma}}}}.
	$
	For the chosen Carreau-Yasuda viscosity model  the dimensional form of the apparent fluid viscosity is related to the second invariant of the strain-rate tensor  $\left( II_{\text{SR}} = \sqrt{\boldsymbol{\overline{\overline{\dot{\gamma}}}}\textbf{:} \boldsymbol{\overline{{\overline {\dot{\gamma}}}}}} \right)  $ by the expression  \citep{Yasuda1981, Bird1987} 
	\begin{equation}\label{eq:tau-CY-dim}
		\widetilde {\mu}(\widetilde{II}_{\text{SR}})= \widetilde {\mu}_\infty + (\widetilde {\mu}_0 - \widetilde {\mu}_\infty) \left( 1+ 2 \widetilde{\lambda}_t^2 \widetilde{II}_{\text{SR}}^2 \right) ^{{(nc-1)}/{2}}. 
	\end{equation}  
	Here $ \widetilde{\mu}_0 $ and $ \widetilde{\mu}_\infty $ are the constant viscosities at zero and infinite shear rates, respectively; $ nc $ is the exponent for generalized Newtonian behaviour ($ nc<1 $ for pseudoplastic (shear thinning) and $ nc>1 $ for dilatant (shear thickening)) and $ \widetilde{\lambda}_t $ is the relaxation time constant signifying the inverse of the shear rate for Newtonian to non-Newtonian transition of fluid rheology. A choice of dimensional reference of $ u_{ref}/a $ for the strain rate, leads to the following dimensionless form for the fluid viscosity:
	\begin{equation}\label{eq:CY}
		\mu(II_{\text{SR}})= \mu_\text{r} + (1 - \mu_\text{r}) \left( 1+ 2 \,\text{Wi}^2 \,II_{\text{SR}}^2 \right) ^{{(nc-1)}/{2}},
	\end{equation}  
	where the Wiessenberg number $(Wi)$ is defined as $ Wi=\widetilde{\lambda}_t u_{ref}/a $ and represents the dimensionless relaxation time constant of the swimming media. The above viscosity model avoids the of infinite shear rate at zero strain-rate for $ nc<1 $. Further depending on the value of the relaxation time, the fluid rheology can resemble either a perfect Newtonian fluid at zero shear rate viscosity $ (\widetilde{\mu}_0) $ (for $ \lambda \to 0$) or a perfect power-law fluid (for $ \lambda \to \infty$).  Thus it becomes a suitable rheological model in predicting shear-dependent fluid viscosity in both in-vivo and in-vitro environments.

	\subsection{Performance assessment of electrocatalytic swimming}
	\label{ssec:eff-calc}
	The primary goal for successful operation of a electrocatalyticaly-driven micromotor is to achieve the maximum possible velocity of the swimmer for a given amount of fuel. However, the optimum utilization of the available fuel source  is a major concern in various situations especially related to the in vivo conditions.  The efficiency of electrocatalytic propulsion is often defined in the form \citep{Wang2013,Kreissl2016}
	\begin{equation}\label{eq:eff-expression}
		\eta_{ep}=\frac{\widetilde{P}_\mathrm{out}}{\widetilde{P}_\mathrm{in}},
	\end{equation}
	where $ \widetilde{P}_\mathrm{out} $ is the energy dissipation in terms of total mechanical power output and $ \widetilde{P}_\mathrm{in} $ is the total chemical power input responsible for the energy of active swimming. 
	
	The chemical power input is calculated by considering the total reaction rate over the swimmer surface and the Gibbs free energy $ (\Delta \widetilde{G}_f) $ required for the dissociation of the fuel $ \mathrm{H_2 O_2} $ into a single $ \mathrm{O_2} $ molecule and water. Hence $ \widetilde{P}_{in} $  takes the form \citep{Kreissl2016}:
	
	\begin{equation}\label{eq:P_in}
		P_\mathrm{in}=  \left( \frac{1}{2}\int_{S}^{}{}  \left |  {\mathbf{\hat n}} \cdot {\mathbf{\tilde J}_S}\right |  dS \right)  \Delta \widetilde{G}_f.
	\end{equation}

	The mechanical power output due to self-electrophoretic motion is calculated by taking into consideration of a similar passive particle which is dragged through the fluid in the same velocity of active-swimming \citep{Sabass2012}. Thus $ \widetilde{P}_\mathrm{out} $ can be expressed as \begin{equation}\label{eq:O/P power}
		\widetilde{P}_\mathrm{out}=\widetilde{F}_\mathrm{D,passive}\, \widetilde{u}_\mathrm{p}.
	\end{equation}
	Here  $ \widetilde{F}_\mathrm{D,passive} $ can be expressed as  $ \widetilde{F}_\mathrm{D,passive}=(6\, \pi \, \widetilde{\mu}_0 \, a \,  \widetilde{u}_\mathrm{p}) \,X_c$, where $ X_c $ is a function of the parameters depicting the deviation from the corresponding  Newtonian fluid with a constant viscosity $ \widetilde{\mu}_0 $. A significant amount of literature exists which attempted to quantify the factor $ X_c $ as a function of Carreau-Yasuda  model parameters $ \beta, Wi$ and $ nc $. 
	\section{Solution methodology}
	\label{sec:sol}
	{The coupled equations of ion transport, Poisson equation, Cauchy momentum equation are first solved using the finite element package COMSOL. Following the earlier investigations by \citet{Kreissl2016} and \citet{Brown2017} we consider an axisymmetric domain and simulate the problem in a reference frame attached to the particle center. Since at the far-stream boundary a stress-free  condition $\left(\left \{-p\,\mathbf I + \mu (\nabla \mathbf{u}+(\nabla \mathbf{u})^T) \right \} \cdot \mathbf{n} =0 \right ) $ is realized, the particle velocity relative to the inertial reference frame, can be obtained by averaging the fluid velocity along this boundary, i.e. $ \widetilde{u}_\mathrm{p} = - \left \langle \mathbf{\widetilde u} \right \rangle \big{|}_B
		$. In order to simulate an unbounded condition it is important to eliminate the effect of surrounding walls on the flow field.}
	
	Before proceeding with the demonstration of various key results, in figure \ref{fig:vabs-vs-chp-vary nc} we show the comparison of present full numerical simulation results for the case $ nc=1 $ with the asymptotic solution of \citet{Sabass2012a} as well as with the experimental results of \citet{Wheat2010}. It is evident that the presently calculated swimmer velocities are in a similar realistic range of the spherical microswimmer experiments by \citet{Wheat2010}. On the other hand, even after using an exact set of parameters the present results differ slightly from the asymptotic theory results. This difference can be attributed to the fact that in their work \citet{Sabass2012a} only considered the leading order terms during the calculation of their singular perturbation analysis with the dimensionless debye length $ \lambda $ as small perturbation parameter. While in stark contrast the full numerical solution procedure adopted in the present work, is not limited to such approximations.
	
	\section{Results and Discussions}
	\label{sec:result}
	In this section we investigate the coupling between the viscous and electrocatalytic parameters on the swimming velocity and the electrocatalytic propulsion efficiency. 
	With a notion to compare our results in the limiting condition of a Newtonian swimming media with those of the \citet{Sabass2012a}, various practical ranges of dimensional parameters are chosen accordingly.  Unless otherwise specified, the Carreau-Yasuda model dimensionless parameters are chosen as $ \mu_\text{r} =0.2$ and $ Wi=10 $. The dimensionless surface reaction coefficients are considered to be related as $ K_r=\kappa_r $ and $ K_o=\kappa_o $ with $ \kappa_r=0.005 $  and $ \kappa_o=0.1$. Also the particle  is having a radius of $ \mathrm{1\;\mu m}$ while the electrocatalytic solution contains ions with bulk ionic concentration 
	$ \widetilde{c}_0=5 \times 10^{-5} $ mol/l. In addition we assume a fixed pH value allowing us to take a fixed value of bulk proton concentration $ (\widetilde{c}_\text{H}^+) $. 
	
	Within the range of practically relevant parameters considered in the present study, the  calculation of the electrocatalytic propulsion efficiency $ \eta_{ep} $ as per section \ref{ssec:eff-calc}, gives a value of $ \eta_{ep} \sim O(10^{-6}-10^{-7}) $. Such a range of $ \eta_{ep} $ is consistent with the experimentally observed efficiency of micromotors by \citet{Wang2013}.

	\subsection{Effects of peroxide concentration}
	\label{ssec:hp-conc-vary}
\begin{figure}
		\centering
		\includegraphics[width=0.85\textwidth]{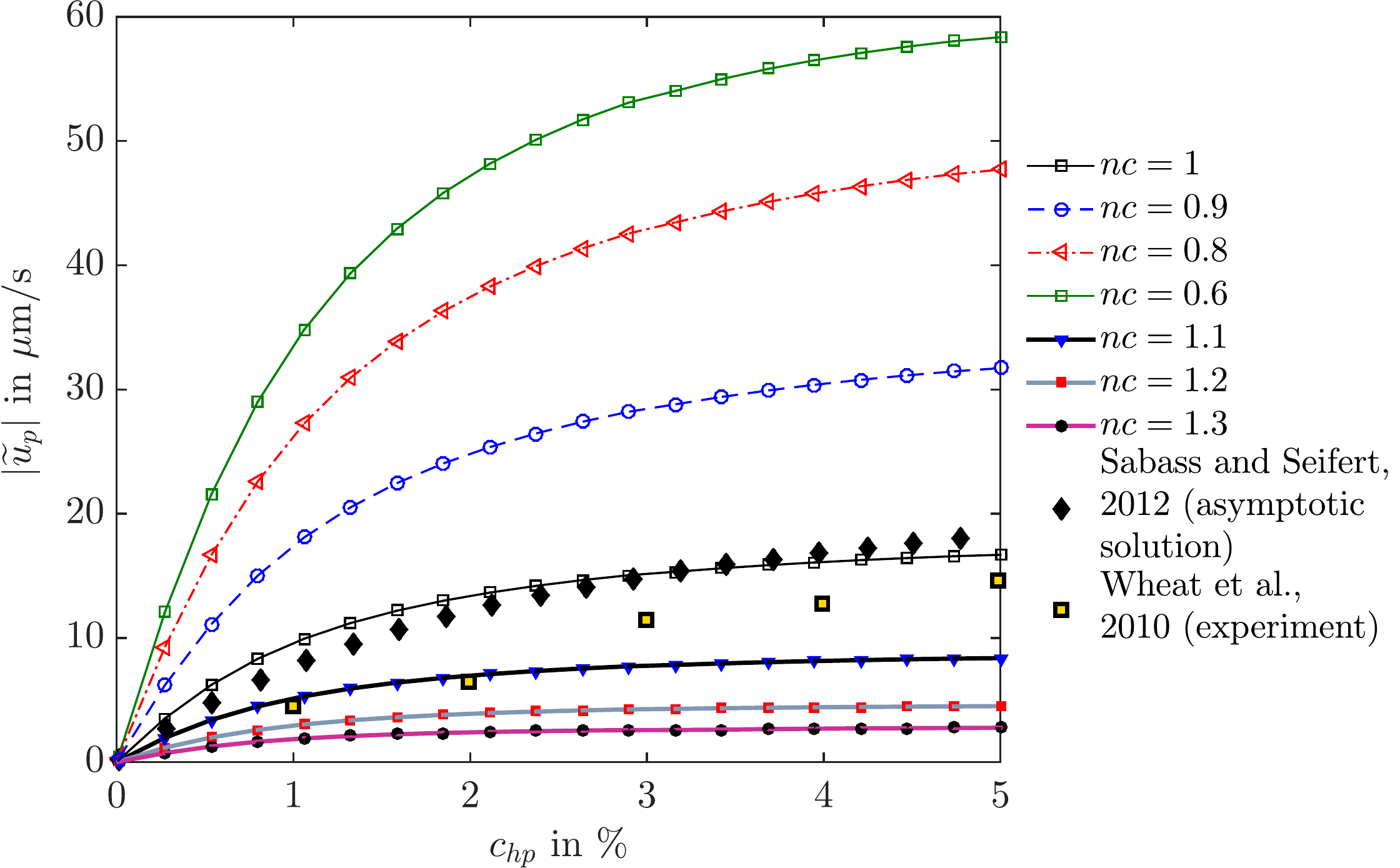}
		\caption{ Swimmer velocity magnitude $ (|\widetilde{u}_\mathrm{p}|) $ vs. hydrogen peroxide concentration $ (c_\text{hp}) $ for different values of the exponent $ nc $. The markers stand for numerically computed data points. The $ nc=1 $ curve for the present work is  compared with the asymptotic solution of \protect{\citet{Sabass2012a}} and experiments of  \protect{\citet{Wheat2010}}.}
		\label{fig:vabs-vs-chp-vary nc}
	\end{figure}

	Figure \ref{fig:vabs-vs-chp-vary nc} depicts how the variations in the dimensional swimmer velocity ($ \widetilde{u}_\mathrm{p} $) is affected by the combined effect of  the fuel concentration $ c_\text{hp} (\%) $ and power-law exponent $ nc $. For a Newtonian swimming media $ \widetilde{u}_\mathrm{p} $ exhibits a non-linear increase with fuel concentration. Thus the particle mobility (defined as $ M_p= \widetilde{u}_\mathrm{p} / c_\text{hp}$) turns out to be  varying with $c_\text{hp}$. In case of a shear thinnig or pseudoplastic fluid the viscosity decreases continuously with shear rate. Various biofluids showing such behaviour are generally constituted of polymer chains dispersed in an aqueous solution. When exposed to a shear rate, these molecular networks face structural disruptions and modifications. As a consequence the viscous resistance to the particle motion  decreases. On the other hand the shear thickening fluids are generally concentrated suspensions showing higher resistance to flow, when sheared. Thus in contrast to the Newtonian media,  with increased shear thinning $ (nc<1) $, $ \widetilde{u}_\mathrm{p} $ experiences a several-fold increase while the reverse happens for the shear thickening fluids $ (nc >1)$. In addition, as the fuel concentration rises, the non-Newtonian swimmer velocity shows an almost constant deviation from the corresponding Newtonian case. This signifies the fact the relative change in mobility $(M_p/M_{pN})$ with varying $ nc $, is very weakly altered with fuel concentration.

	\subsection{Effects of Weissenberg number}
	\label{ssec:Wi-n-vary}
\begin{figure}
		\centering
		\begin{subfigure}[!htbp]{0.45\textwidth}
			\centering
			\hspace{-8ex}
			\includegraphics[width=1.155\textwidth]{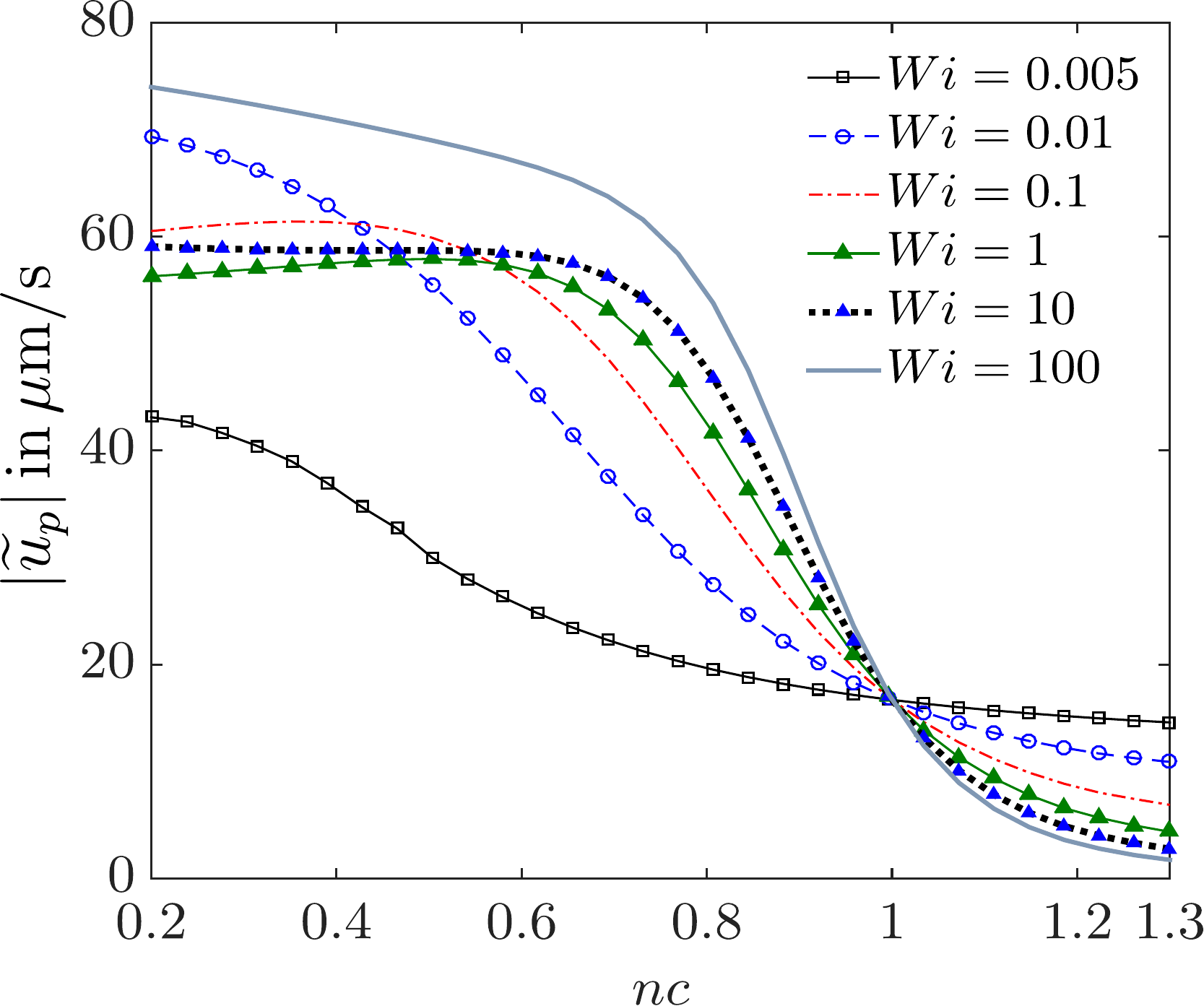}
			\vspace{12ex}
			\caption{}
			\label{fig:vabs-vs-nc-vary-Wi}
		\end{subfigure}
		%	\\[-12ex] 
		\begin{subfigure}[!htbp]{0.45\textwidth}
			\centering
			\includegraphics[width=1.12\textwidth]{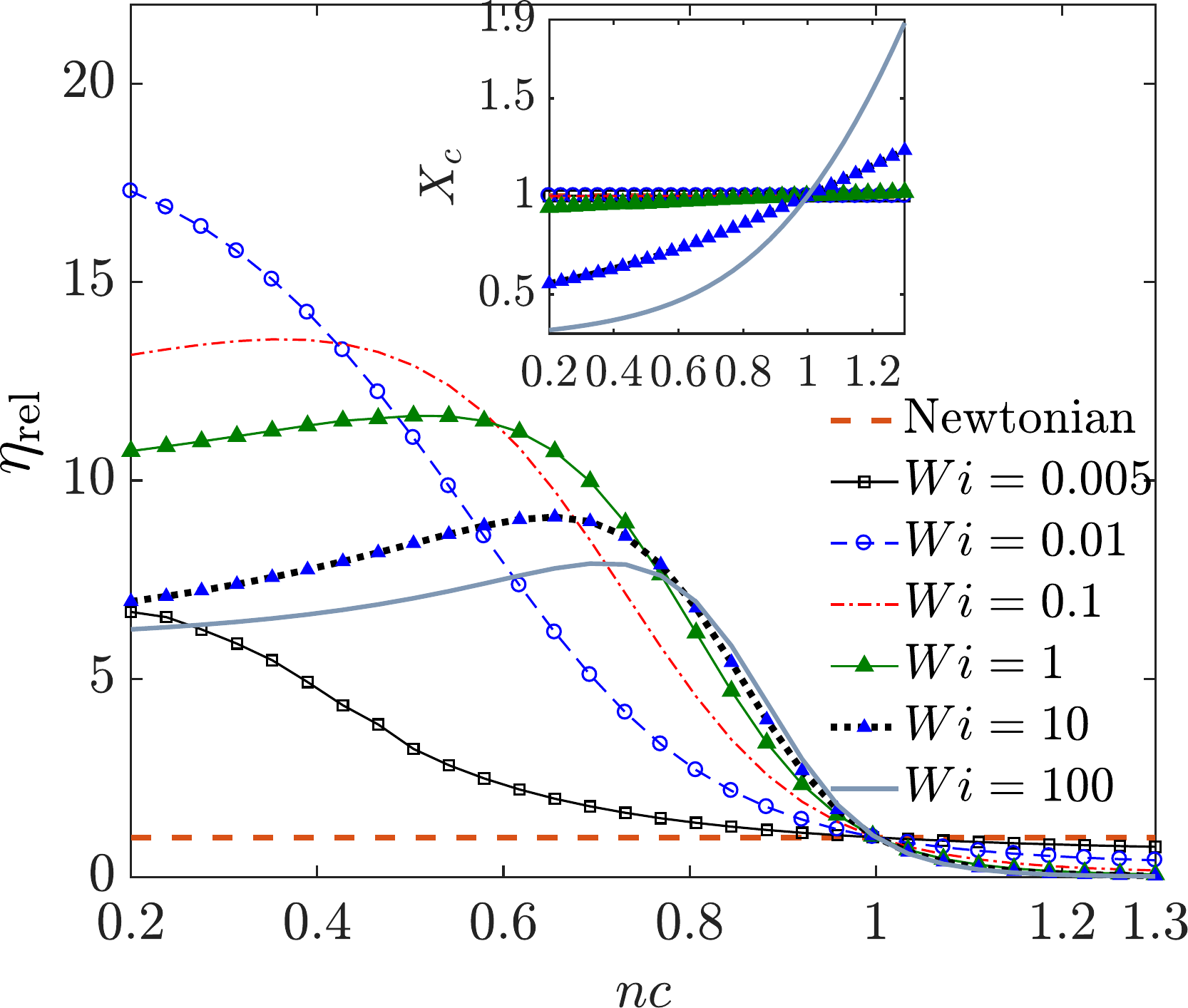}		
			\vspace{8ex}
			\caption{}
			\label{fig:eta-rel-vs-nc-vary-Wi}
		\end{subfigure}
		\caption{  (a) Swimmer velocity magnitude $ (|\widetilde{u}_\mathrm{p}|) $ vs. exponent $ nc $ for different values of the Weissenberg number $(Wi)$. (b) Relative swimmer efficiency $ \eta_\mathrm{rel} $ vs. power law exponent $ nc $ for different values of the Weissenberg number $(Wi)$. In both the subplots  $ c_\text{hp}=5 (\%) $ is taken.} 
		\label{fig:up-vs-chp and nc}
	\end{figure}

	Weissenberg number $(Wi)$ physically signifies the dimensionless relaxation time of the fluid to adjust itself against an external shear. When an external flow is imposed, depending upon the inherent fluid-structural properties, the shear thinning or shear thickening nature is either intensified (for $ Wi \gg 1$) or suppressed (for $ Wi \ll 1$).  Figure \ref{fig:vabs-vs-nc-vary-Wi} depicts that for shear thickening $ (nc>1) $ and moderately shear thinning $ (0.6 \lesssim nc<1) $ rheology, an increase in $ Wi $ leads to increasing deviations in swimmer velocity from the   Newtonian case. However for $nc \gtrsim 0.6 $ the effects of power law index $ (nc) $ and dimensionless relaxation time $ (Wi) $ compete with each other in deciding the fluid flow pattern and in turn the drag force on the particle. As a result the particle velocity does not increase in  a monotonic fashion with $ Wi $. Moreover, for intermediate values of $ Wi $(= 0.1,1 and 10), $ |\widetilde{u}_\mathrm{p}| $ reaches a nearly saturating condition beyond $ nc \lesssim 0.6$. Thus the Weissenberg acts in a highly non-trivial and complex manner in gaining a greater velocity with increasing pseudoplasticity of  the surrounding fluid.  
	
	Figure \ref{fig:eta-rel-vs-nc-vary-Wi} shows while for shear thickening medium the efficiency is even lower than the Newtonian case, a significant gain is achieved for the shear thinning case. For $0.8 \lesssim nc<1$, an increase in $ Wi $ causes an evident increase in $ \eta_\mathrm{rel} $ before reaching the maximum attainable efficiency for purely power-law situation (for $ Wi\sim100$). 
	However the scenario is completely different as $ nc $
	becomes $ \lesssim 0.8 $. With a continuous decrease in $ nc $, the $ \eta_\mathrm{rel} $ reaches a maximum value and thereafter reduces again. Also such a behaviour is strongly dependent on the  value of $ Wi$ and the existence of an optimality condition occurs only when $ Wi$ becomes moderate or high ($ Wi=0.1,1,10 \text{\,\,and\,\,} 100$). 
	This striking occurrence can be described with due consideration of the simultaneous modifications in swimmer velocity and  reduction in the drag correction factor $ X_c $ at lower values of $ nc $. In the inset of figure \ref{fig:eta-rel-vs-nc-vary-Wi}, the corresponding variations in the drag correction factor $ X_c $ establish the physical phenomenon that the cost of dragging an equivalent passive particle in the form of viscous dissipation, gets reduced with decreasing $ nc $. Referring back to figure \ref{fig:vabs-vs-nc-vary-Wi}, for these values of $ Wi $ the swimmer velocity showed a trend of varying very weakly beyond a certain reduction in $nc$. On the other hand the drag correction factor shows a sharp decline even in these low value of $nc$.  Hence the behaviour of the two effects becomes competitive in this region and an optimality condition results. 
	
	Another important aspect of the relative efficiency variation is the shift of the optimum points $ (nc_\text{opt}) $ towards increasing $ nc $ with a rise in $ Wi$. This can be attributed to the physical consequences that both swimmer velocity and the passive drag dissipation, are critically altered by changes in $ Wi$, which finally decides the location of the $ nc_\text{opt} $.

	\subsection{Effects of the reaction constants}
	\label{ss:k1-k2-effect}
\begin{figure}[!htbp]
		\centering
		\begin{subfigure}[!htbp]{0.45\textwidth}
			\centering
			\hspace{-8ex}
			\vspace{3ex}
			\includegraphics[width=1.12\textwidth]{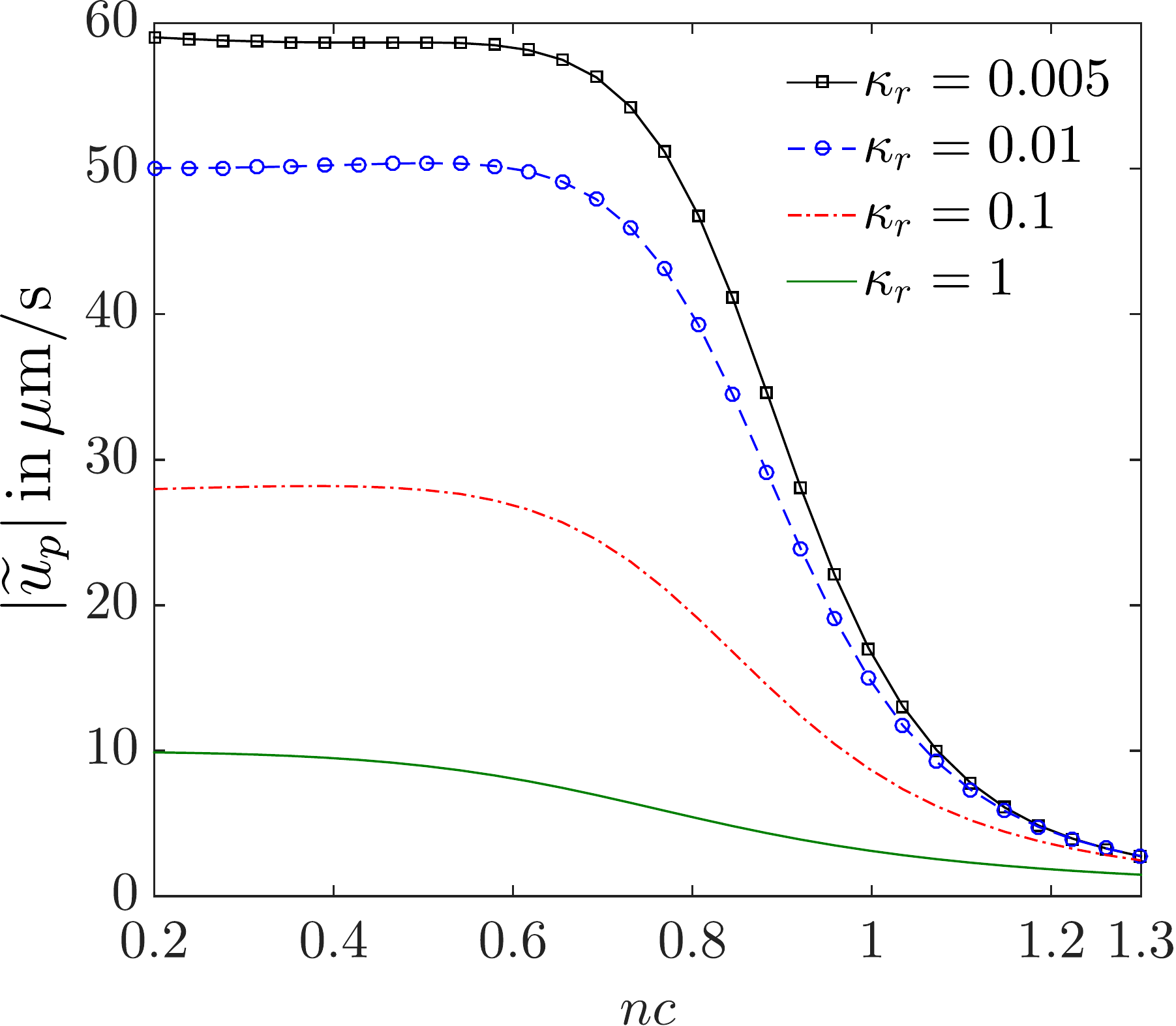}		
			\vspace{8ex}
			\caption{}
			\label{fig:V_abs_vs_NC_vary_kappa1_R}
		\end{subfigure}
		\begin{subfigure}[!htbp]{0.45\textwidth}
			\centering
			\includegraphics[width=1.08\textwidth]{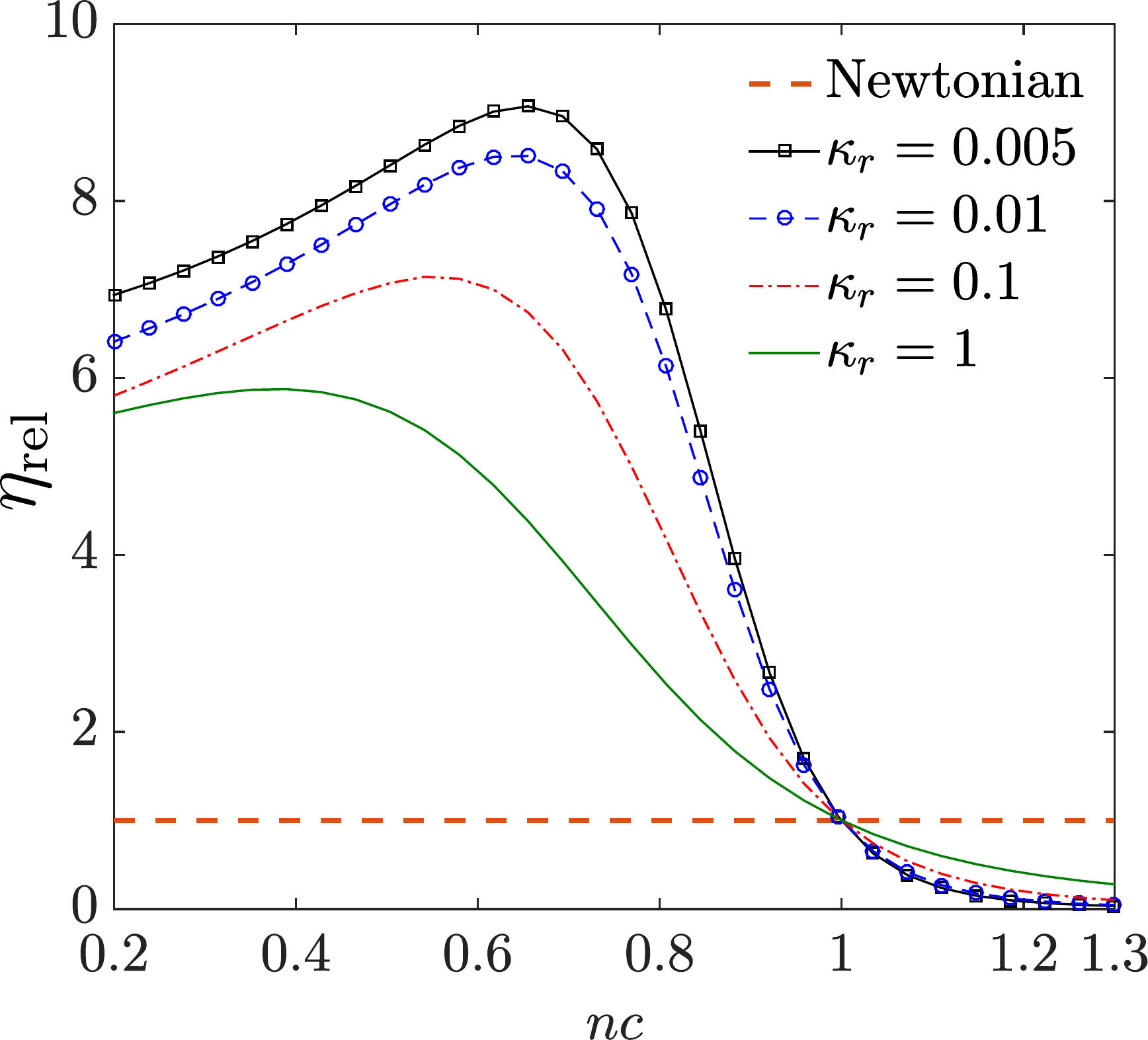}
			\vspace{8ex}
			\caption{}
			\label{fig:EFF_REL_vs_NC_vary_kappa1_R}
		\end{subfigure}
		%	\\[-12ex] 
		\begin{subfigure}[!htbp]{0.45\textwidth}
			\centering
			\hspace{-6ex}
			\vspace{1.7ex}
			\includegraphics[width=1.12\textwidth]{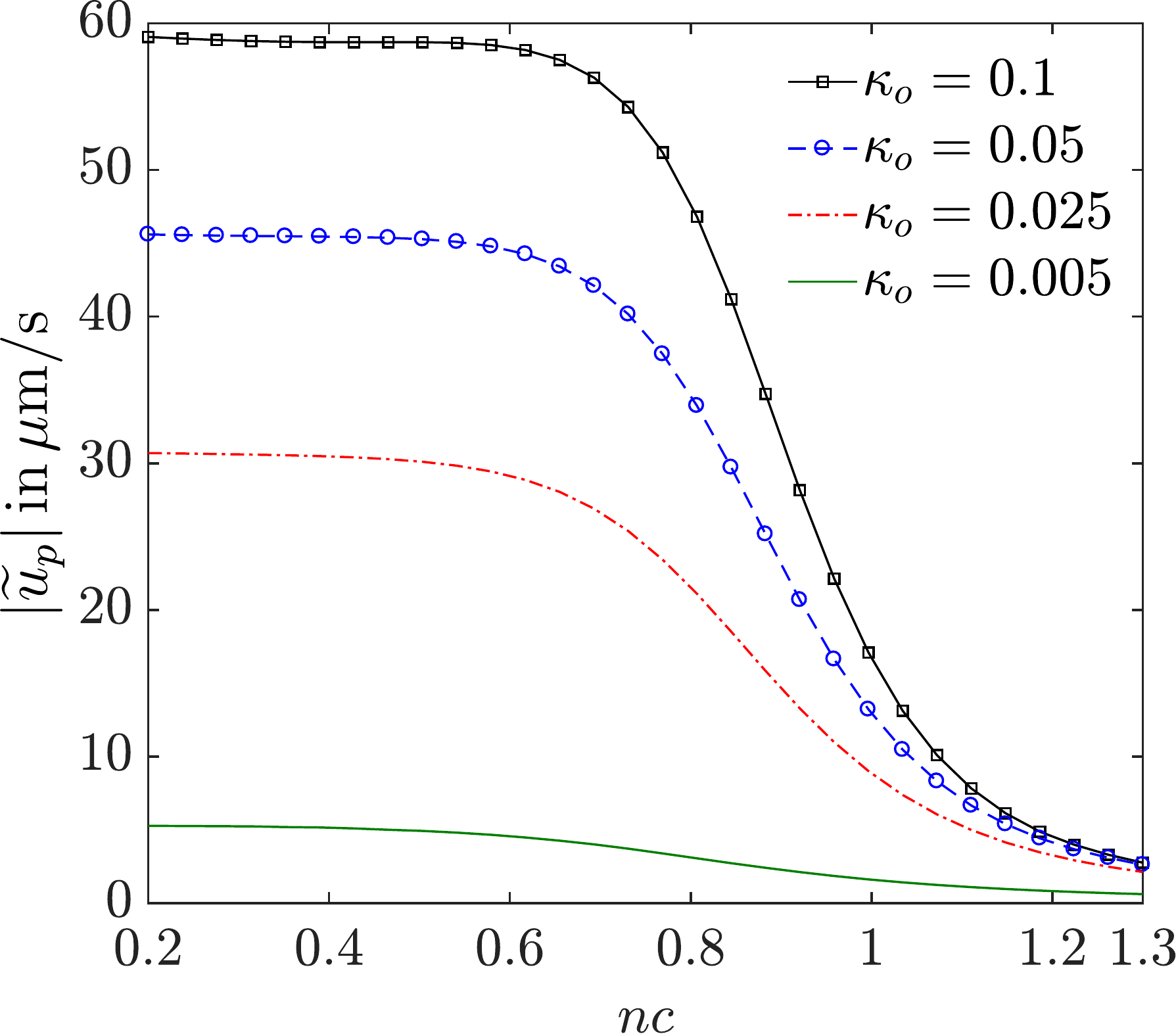}		
			\vspace{6.5ex}
			\caption{}
			\label{fig:V_ABS_vs_NC_vary_kappa2_O}
		\end{subfigure}
		\quad 
		\begin{subfigure}[!htbp]{0.45\textwidth}
			\centering
			%			\vspace{1.8ex}
			\includegraphics[width=1.1\textwidth]{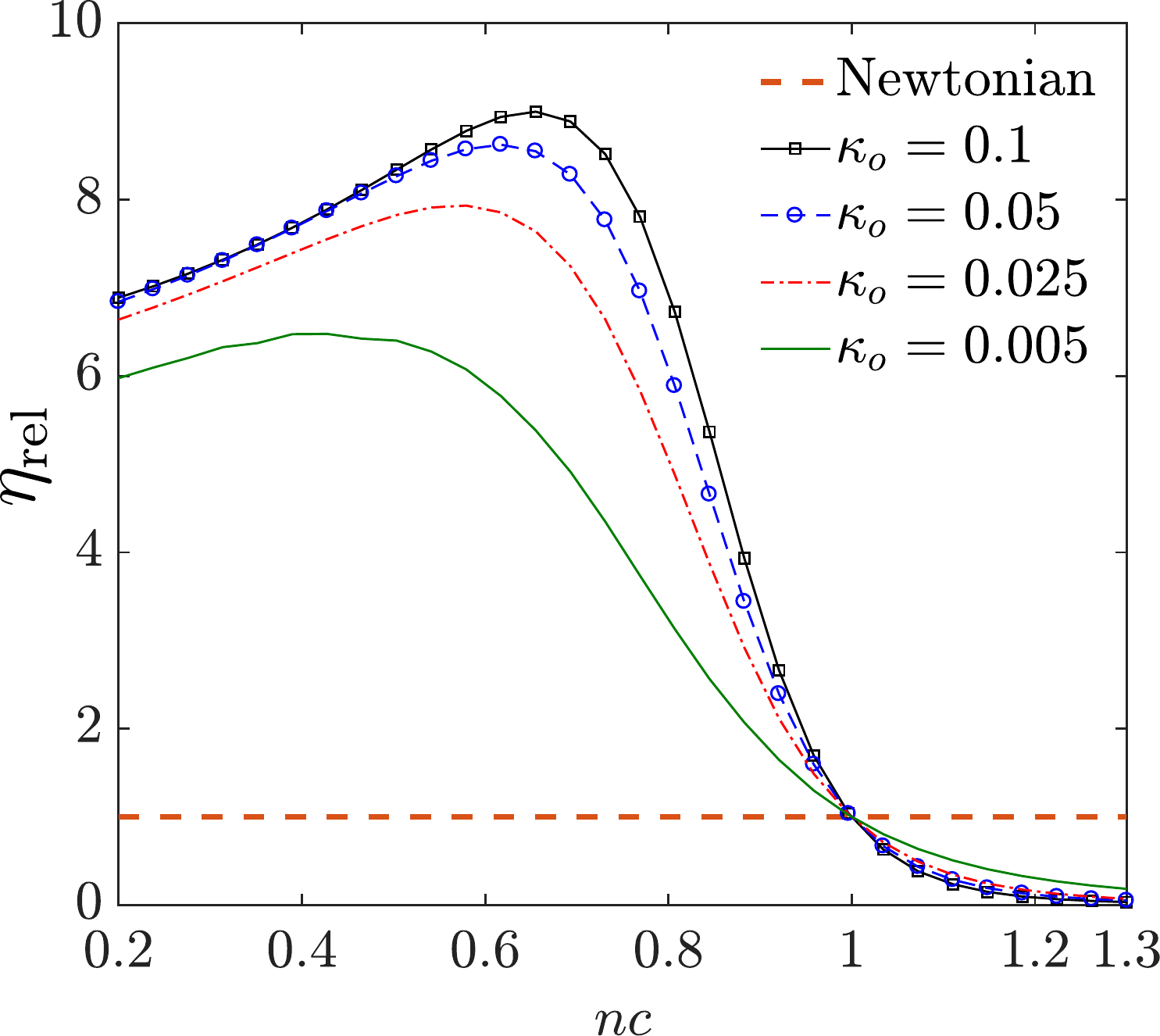}
			\vspace{8ex}
			\caption{}
			\label{fig:EFF_REL_vs_NC_vary_kappa2_O}
		\end{subfigure}
		\caption{(a) Swimmer velocity magnitude $ (|\widetilde{u}_\mathrm{p}|) $ vs. exponent $ nc $ for different values of the reduction constant $(\kappa_r)$. (b) Relative swimmer efficiency $ (\eta_\mathrm{rel}) $ vs. exponent $ nc $ for different values of the reduction constant $(\kappa_r)$. (c) Swimmer velocity magnitude $ (|\widetilde{u}_\mathrm{p}|) $ vs. exponent $ nc $ for different values of the oxidation constant $(\kappa_o)$. (d) Relative swimmer efficiency $(\eta_\mathrm{rel})$ vs. exponent $ nc $ for different values of the oxidation constant  $(\kappa_o)$. In all the subplots $ c_\text{hp}=5 (\%) $ is taken.} 
		\label{fig:reduction-oxidation-consts}
	\end{figure}

	%{\color{BurntOrange}
	In figure \ref{fig:reduction-oxidation-consts} we demonstrate the coupled effects of reaction constants (reduction constant, $ \kappa_r $ and oxidation constant,  $ \kappa_o $) and fluid rheology parameters on swimmer velocity and relative efficiency of the electrocatalytic propulsion phenomenon. Comparing figures \ref{fig:V_abs_vs_NC_vary_kappa1_R} and 		\ref{fig:V_ABS_vs_NC_vary_kappa2_O}, it can be observed that the increasing reduction constant causes a delay in reaching a constant velocity for low values of $ nc $, while the similar effect is showed by decreasing oxidation constant values. In a similar way a careful comparative study of figures \ref{fig:EFF_REL_vs_NC_vary_kappa1_R} and 	\ref{fig:EFF_REL_vs_NC_vary_kappa2_O} reveals that with either an increase in $ \kappa_r $ or decrease in $ \kappa_o$, the relative gain in efficiency shows an increasing trend. It is also interesting to find that the optimality condition $ nc_\text{opt}$ is also hugely altered by a varying degree of the said reaction constants. For analyzing the physical mechanism behind such non-trivial interplay, we delve deeper to investigate the effects of surface kinetics on the fluid flow behaviour. The detailed surface kinetics, as opposed to a prescribed surface potential or surface charge models, is a function of these reaction constants. They ($\kappa_r \, \text{and} \,  \kappa_o $) not only influence the absolute average cation flux $(|\widetilde{\alpha}|_\mathrm{av}) $ but also decide the effective surface potential $ (V_S) $  on the particle surface for the chosen set of parameters.
	The magnitude of surface potential shows decreasing and increasing trends trends with variations in $ \kappa_r $ and $ \kappa_o$,  respectively (not shown for brevity). This consequently results in significant modifications in the concentration distributions of both the active species and the salt ions. Finally the electrical body force acting on the adjacent swimming medium, is altered severely. A varying degree of electrical body force interacts differently with the shear-rate dependent viscous response of the swimming medium. In effect the drag force experienced by the particle is also modified thereby causing evident changes in the swimming velocity $ (\widetilde{u}_\mathrm{p}) $ and the rheology-induced modifications in the electrocatalytic efficiency $ (\eta_\text{rel}) $.
	%}
	
\begin{figure}
		\centering
		\begin{subfigure}[!htbp]{0.45\textwidth}
			\centering
			\hspace{-8ex}
			\includegraphics[width=1.155\textwidth]{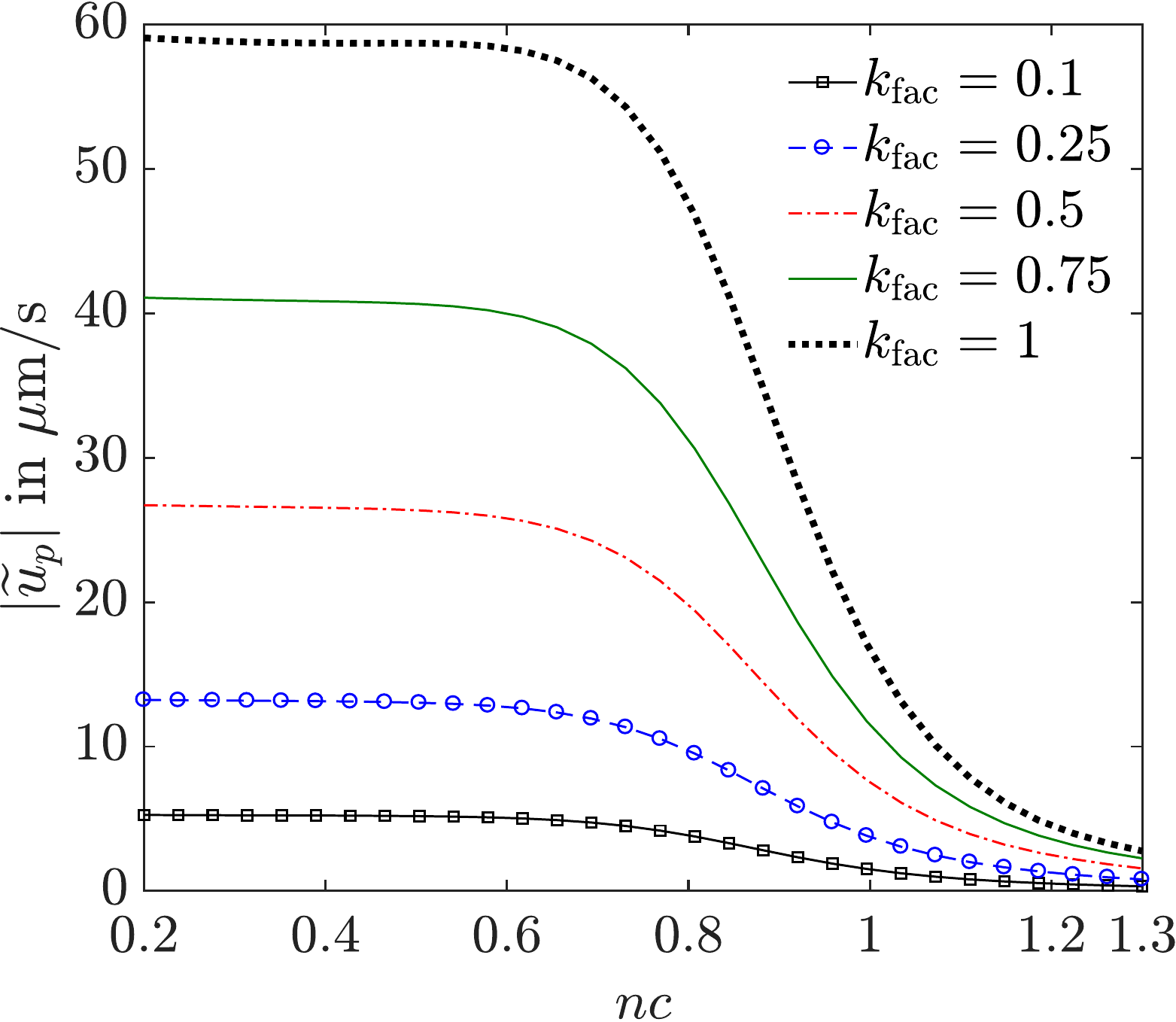}
			\vspace{12ex}
			\caption{}
			\label{fig:V_ABS_vs_NC_vary_K-factor}
		\end{subfigure}
		%	\\[-12ex] 
		\begin{subfigure}[!htbp]{0.45\textwidth}
			\centering
			\vspace{-0.75ex}
			\includegraphics[width=1.152\textwidth]{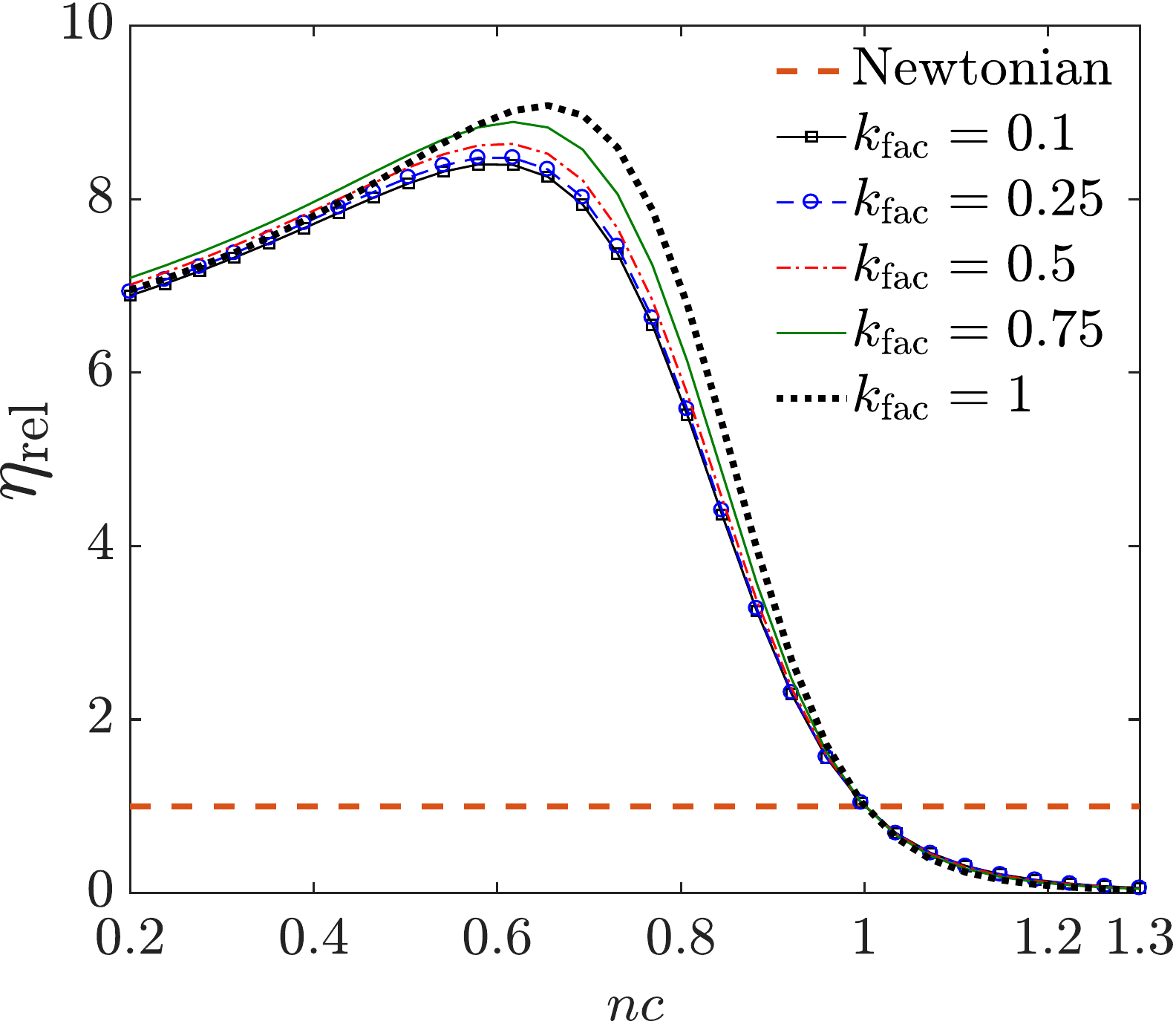}		
			\vspace{8.2ex}
			\caption{}
			\label{fig:EFF_REL_vs_NC_vary_K-factor}
		\end{subfigure}
		\caption{(a) Swimmer velocity magnitude $ (|\widetilde{u}_\mathrm{p}|) $ vs. power law exponent $ nc $ for different values of the relation factor for the reaction constants,  $k_\mathrm{fac} = K_r/\kappa_r=K_o/\kappa_o $. (b) Similar variation of the relative swimmer efficiency $ (\eta_\mathrm{rel})$. In both the subplots (a) and (b),  $ c_\text{hp}=5 (\%) $ is taken.}
		\label{fig:up-eff-k-fac}
	\end{figure}

	In order to quantify the effects of heterogeneity in the distribution of oxidation and reduction sites across the particle surface, we define the following relation factor for different reaction constants, given as  $k_\mathrm{fac} = K_r/\kappa_r=K_o/\kappa_o $. Physically the low values of this factor (e.g. $ k_\mathrm{fac}=0.1$) signify weak variation in reactivity while a high value (e.g. $ k_\mathrm{fac}=1$) stands for strong variation of the same quantity \citep{Sabass2012a}. Figure \ref{fig:V_ABS_vs_NC_vary_K-factor} shows that similar to the previously discussed effects of $ \kappa_r $ and $ \kappa_o$, the  swimmer velocity gets enhanced with increasing asymmetry in the reaction pattern (i.e. rising value of $ k_\mathrm{fac}$). In contrast, as depicted in figure \ref{fig:EFF_REL_vs_NC_vary_K-factor}, only a marginal change in the relative efficiency is observed. However optimum efficiency point $ (nc_\mathrm{opt}) $ shifts towards $nc \approx	 0.7$ as $ k_\mathrm{fac} \to 1$. These observed changes due to $ k_\text{fac} $ can again be attributed to corresponding alterations in the surface potential and distortions in the concentration patterns.

	\subsection{Effects of bulk salt ion concentration}
	\label{ss:n0-effect}
\begin{figure}[!htbp]
		\centering
		\begin{subfigure}[!htbp]{0.45\textwidth}
			\centering
			\hspace{-10ex}
			\includegraphics[width=1.22\textwidth]{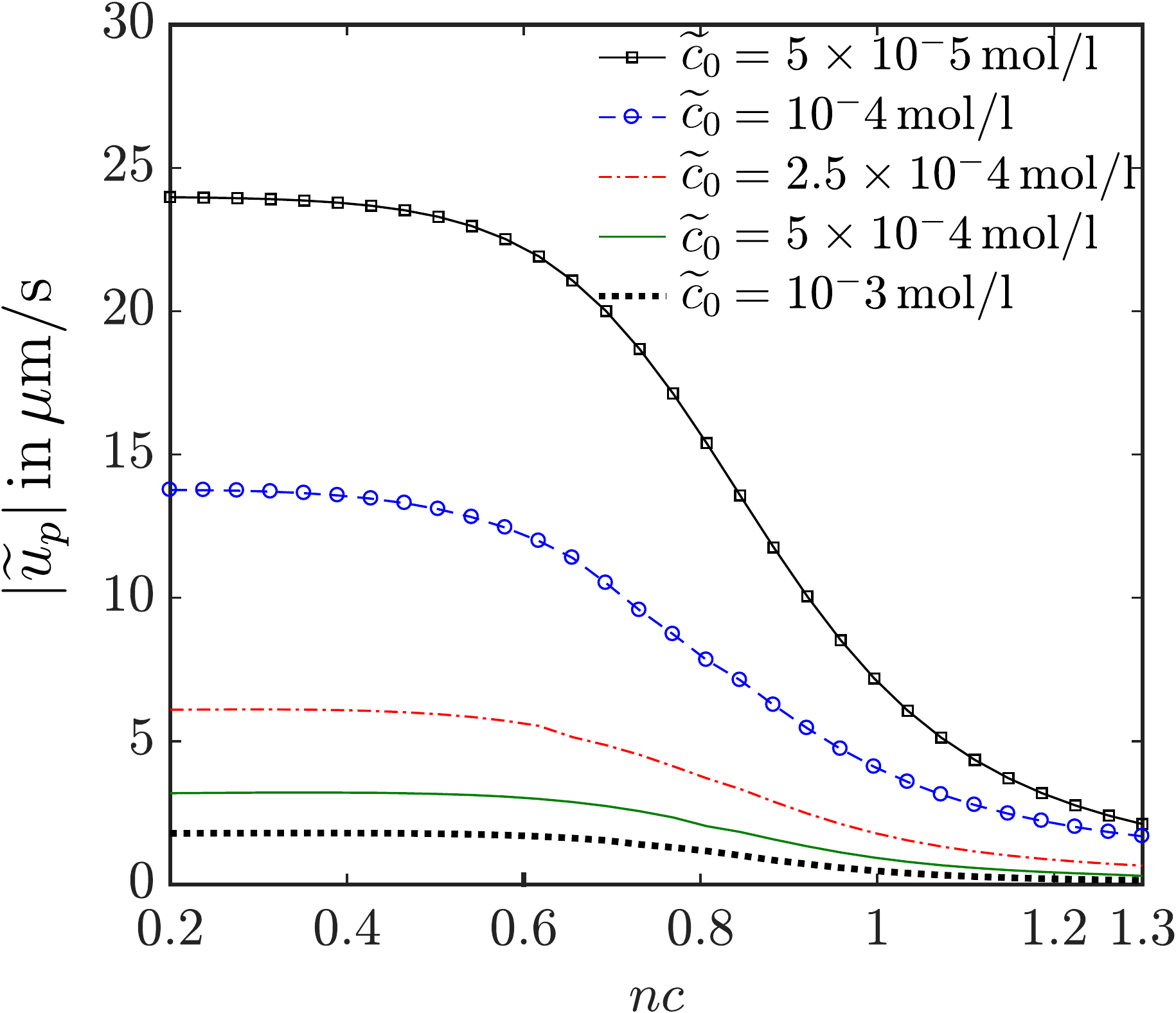}		
			\vspace{8ex}
			\caption{}
			\label{fig:vel-c0-effect}
		\end{subfigure}
		\begin{subfigure}[!htbp]{0.45\textwidth}
			\centering
			\includegraphics[width=1.175\textwidth]{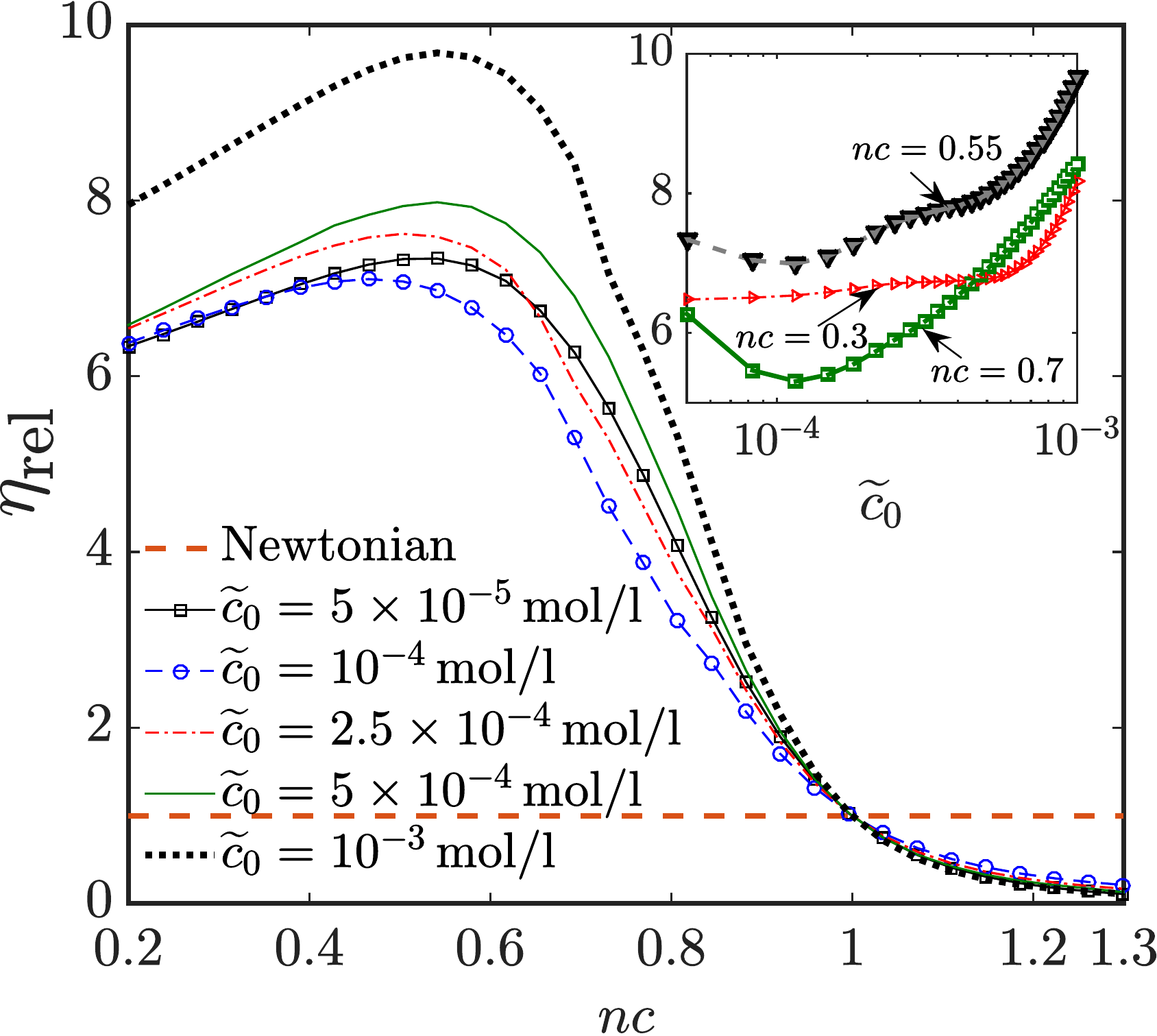}
			\vspace{8ex}
			\caption{}
			\label{fig:eff-c0-effect}
		\end{subfigure}
		\caption{(a) Swimmer velocity magnitude $ (|\widetilde{u}_\mathrm{p}|) $ vs. exponent $ nc $ for different values of the bulk salt ion concentration $(\widetilde{c}_0)$. (b) Similar variation of the relative swimmer efficiency $(\eta_\mathrm{rel})$. In both the subplots (a) and (b),  $ c_\text{hp}=3.7 (\%) $ is taken. In the inset of subplot (b), the variation of $ \eta_\mathrm{rel}$ with $ \widetilde{c}_0 $ is shown for 3 different power law exponents $ (nc) $.} 
		\label{fig:c0-effect}
	\end{figure}

	The reductionin swimmer velocity with increase in the added salt concentration, as portrayed in figure \ref{fig:vel-c0-effect}, was also previously observed by Brown and co-workers \citep{Brown2017,Brown2014,Brown2016} for  both Pt - polystyrene (PS) and Au-Pt janus swimmers in Newtonian medium. 
	In the present context of non-Newtonian medium, the above trend remains unaltered, while the amount of velocity reduction becomes a  strong function of the power law exponent $ (nc)$. As the fluid becomes increasing pseudoplastic $ (nc<1) $ starting from a dilatant $ (nc>1) $ one, the bulk-concentration causes the  velocity to decrease more significantly. 
	
	Figure  \ref{fig:eff-c0-effect} demonstrates the effect of bulk salt concentration on the relative propulsion  efficiency of the swimmer $ (\eta_\text{rel}) $. Unlike $ \widetilde{u}_\text{p}$, the $ \eta_\text{rel} $ vs. $ nc $ profiles do not follow a fixed trend. Both the non-Newtonian medium particle velocity $ (\widetilde{u}_\text{p}) $ and  velocity in Newtonian medium $ (\widetilde{u}_\text{pN}) $, are altered with changes in $ \widetilde{c}_0$. The actual gain or loss in efficiency due to non-Newtonian rheology, critically depends on $ \widetilde{c}_0 $ and $ nc$. The figure also shows that for a high value of $ \widetilde{c}_0$, the efficiency is the highest as compared to other concentration values. However, employing such a high salt concentration may be hampered by the practical requirement high swimmer velocity. In the inset of the same figure we show the existence of minimum efficiency points with varying $ \widetilde{c}_0$, for some of the $ nc $ values.  Thus for the practical circumstances, a balance between the requirement of an increased swimmer speed and optimum utilization of available fuel resources, will determine the actual salt concentration to be employed.

	\subsection{Effects of particle diameter}
	\label{ss:d-effect}
	\begin{figure}[!htbp]
		\centering
		\begin{subfigure}[!htbp]{0.45\textwidth}
			\centering
			\hspace{-6.5ex}
			\vspace{1.5ex}
			\includegraphics[width=1.125\textwidth]{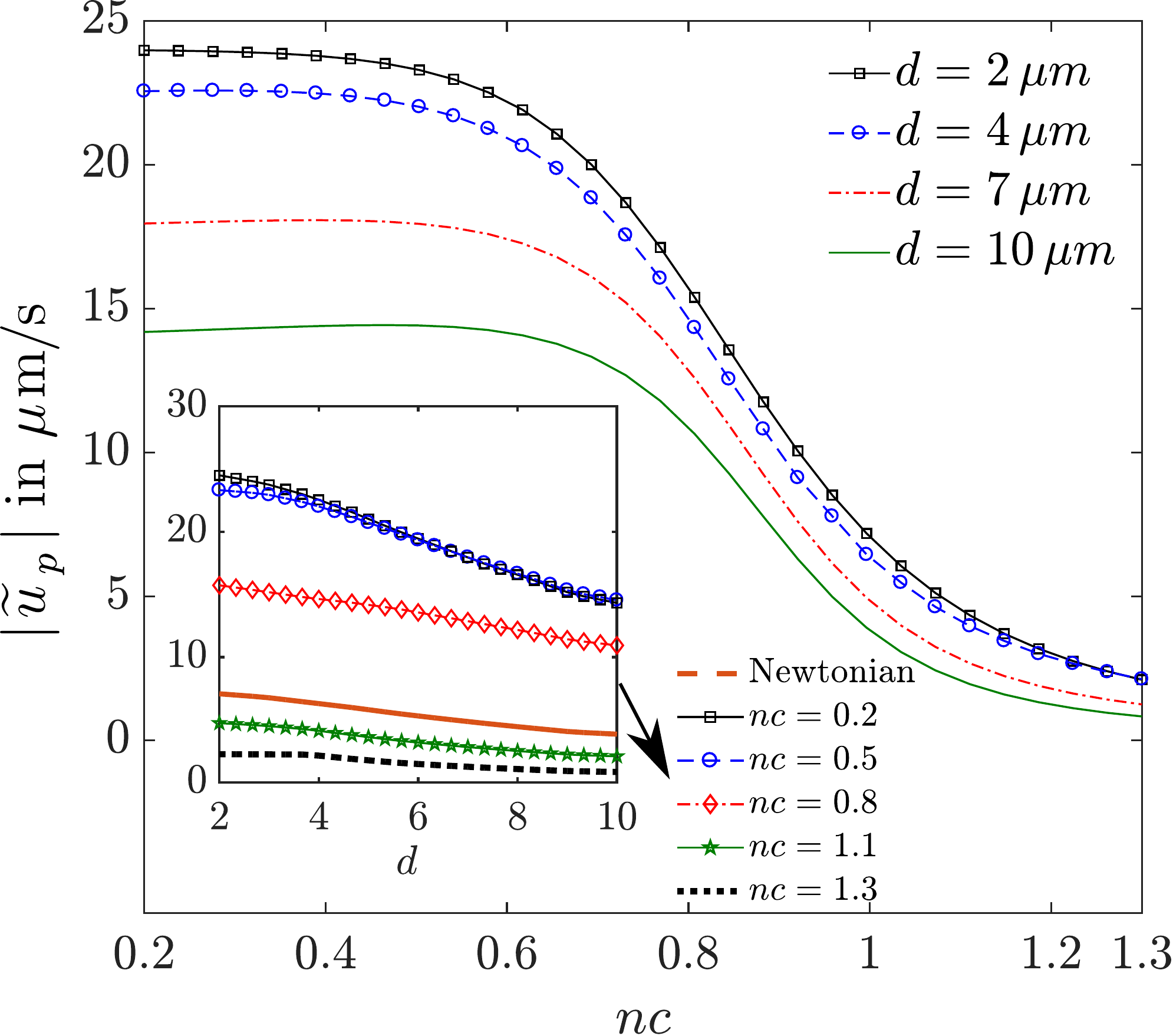}		
			\vspace{6.7ex}
			\caption{}
			\label{fig:Vel_ABS_vs_NC_vary_Length_inset}
		\end{subfigure}
		\begin{subfigure}[!htbp]{0.45\textwidth}
			\centering
			\includegraphics[width=1.11\textwidth]{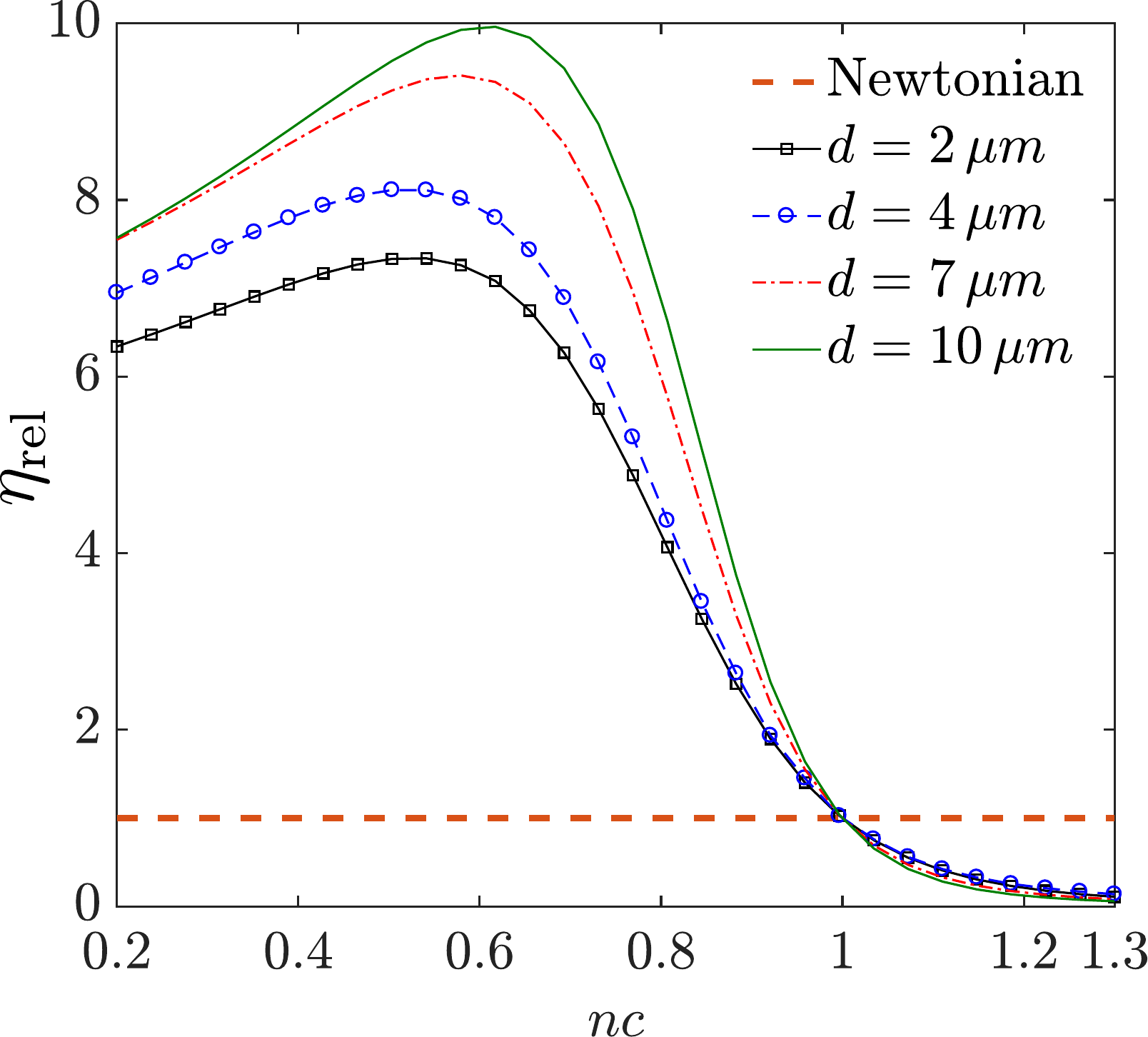}
			\vspace{8ex}
			\caption{}
			\label{fig:EFF_REL_vs_NC_vary_Length}
		\end{subfigure}
		\caption{(a) Swimmer velocity magnitude $ (|\widetilde{u}_\mathrm{p}|) $ vs. exponent $ nc $ for different values of the particle diameter $(d)$. (b) Similar variation of the relative swimmer efficiency $ (\eta_\mathrm{rel})$. In the inset of subplot (a) $|\widetilde{u}_\mathrm{p}|$ vs. $d$ is shown for different values of $ nc $. In both the subplots (a) and (b),  $ c_\text{hp}=3.7 (\%) $ is taken.} 
		\label{fig:particle-size}
	\end{figure}
	Here we investigate the effects of particle size on the swimmer velocity and relative efficiency for different rheological conditions. Figure \ref{fig:Vel_ABS_vs_NC_vary_Length_inset} describes that as the power law exponent $ nc $ decreases, more and more velocity gain is achieved with reduction in particle diameter.  To obtain a clear view of this effect, in the inset of the same figure, the variation of the swimmer velocity with increasing diameter is shown. It shows that for all the chosen $ nc $ values, the swimming velocity decreases approximately linearly with increasing particle diameter. However, with an increased shear thinning or thickening behaviour, the slope of the linear dependency is altered. Similar trend for a Newtonian swimming medium was observed in earlier experiments by \citet{Wheat2010}. However, they attributed the contrast between the dimensional scaling analysis and the experimental observation, to the probable viscous and/or electrokinetic interaction of the swimmer with the nearby solid substrates \citep{Wheat2010,Moran2017}. Also a contradictory increase in particle velocity with diameter was previously predicted based on either a constant surface potential (or constant surface charge) \citep{Brown2017} or dimensional scaling argument with a thin electrical double layer approximation \citep{Golestanian2007}. Quite differently, in the present numerical modeling we have properly taken into account the effects of particle diameter, peroxide concentration and proton concentration on the zeta potential  through a Michaelis-Menten  like surface reaction kinetics.  Functional dependency of both the cation flux and zeta potential on the particle diameter, influences the electrical body force for the movement of the adjacent fluid. This in turn governs the drag force on the particle. As an overall consequence an increase in the particle velocity is observed with deceasing diameter, even in the absence of any confinement effects. 
	
	Although the swimmer velocity shows a decreasing trend when the swimmer diameter is increased, the concern of efficient fuel utilization is best satisfied for bigger swimmer size (shown in figure   \ref{fig:EFF_REL_vs_NC_vary_Length}). Additionally the optimum efficiency power law index $ (nc_\text{opt})  $ for shear thinning swimming medium, becomes larger with increasing particle diameter.

	\section{Conclusions}
	\label{sec:Conclusions}
	In the present study, we theoretically modeled the electrocatalytic swimming of a bimetallic swimmer in a non-Newtonian medium.  The derived equations were then solved numerically using a finite element framework.  We described the effects of various electrocatalytic parameters such as the fuel concentration $ (c_\mathrm{hp} (\%)) $, the reduction and oxidation rate constants ($ \kappa_r  $ and $ \kappa_o $, respectively), the degree of variation of reaction constants $ (k_\mathrm{fac})$, bulk salt ion concentration $(\widetilde{c}_0)  $ and the particle size $ (d) $, on the rheological alterations (quantified by Carreau-Yasuda model parameters $ nc $ and $ Wi $) on the swimming characteristics. Apart from the swimming velocity, the variations in relative electrocatalytic propulsion efficiency $ (\eta_\mathrm{rel}) $ were also discussed.
	The important observations can be summarized as follows:
	
	\begin{enumerate}[label=(\roman*),leftmargin=0pt,itemindent=3em]
		\item The swimming velocity experiences a signifant enhancement as the surrounding fluid becomes more and more shear thinning $ (nc<) $ in nature, while the opposite trend is observed with increasing shear thickening effect $ (nc>1)$. Moreover, the lesser the infinite shear stress viscosity $ (\widetilde{\mu}_\infty) $ becomes as compared to the zero shear stress viscosity $ (\widetilde{\mu}_0) $, the more the augmentation of $ \widetilde{u}_\mathrm{p}$ becomes. 
		
		\item 
		As the fluid rheology shifts more quickly from the Newtonian to non-Newtonian nature (quantified with Weissenberg number, $ Wi $), the relative electrocatalytic efficiency $ (\eta_\mathrm{rel})$ becomes maximum for some specific values of the power law index $ (nc_\mathrm{opt})$. With $ Wi $ rising, while the maximum efficiency  becomes lower, the optimum power law indices locate more towards $ nc \sim 1$. This behaviour of $ \eta_\mathrm{rel}$ can be attributed to differently altered swimmer velocity $ ({u}_\mathrm{p}) $ and drag force correction factor  $ (X_c) $ due to complex rheology. 
		\item 
		Rheology-driven modifications in the electrocatalytic efficiency $ (\eta_\text{rel}) $, are highly coupled with the reaction factors $ (\kappa_r, \kappa_o$ and $ k_\mathrm{fac}) $. In addition to altering the cation-flux at the swimmer surface, these parameters also modify the surface potential and  charge distribution. As a resulting effect, the oxidation and reduction constants influence the swimmer velocity in a different fashion. The optimum efficiency conditions  $ (nc_\mathrm{opt},\eta_\mathrm{rel,max})$ are also decided by a complex interplay between the reaction constants and rheology.
		
		\item   Increased bulk salt ion concentration $ \widetilde{c}_0$ always causes a loss in swimmer velocity. However the gain in efficiency is not always favourable with decreased values of $ \widetilde{c}_0$. Here the rheology-induced velocity field modifications compete with the altered body force on the fluid due to changes in salt concentration. Hence based on the requirement of the most important output criteria between high speed in swimming and efficient utilization of fuel resources, a suitable salt concentration has to be employed.
		
		\item  Our study reveals the importance of a detail reaction kinetics model in order to describe the experimentally observed velocity reduction with increased swimmer diameter. Although the swimmer velocity gets reduced, the efficiency of the operation gets enhanced with the particle diameter increasing towards a value of 10 $ \mu $m.
		
	\end{enumerate}
	
	For the parametric space chosen for demonstration of the results, as high as $ \approx $ 10 times enhancement in the electrocatalytic propulsion efficiency is observed.  This provides a new avenue in  selecting the optimum operating conditions of self-electrophoretic swimming where a significantly high efficiency can be achieved without meeting the requirement of a complicated fabrication technique to generate pre-determined, localized reactive sites. In this respect the present paper stands as a precursor to detail experimental verifications of predicted favourable operating conditions for the auto-electrophoretic swimming of synthetic microswimmers in non-Newtonian medium.

	\bigskip
	%\newpage
	\FloatBarrier
	\bibliographystyle{jfm}
%	\bibliography{Janus_new_only_this}

\end{document}